\documentclass[journal=nalefd,manuscript=article]{achemso}

\usepackage{xcolor}
\usepackage[version=3]{mhchem}
\usepackage{nameref}
\usepackage[colorlinks=true, linkcolor=black, citecolor=black, urlcolor=black]{hyperref}

\title{Tunable magnon emission from a nano-optomagnet}

\author{A. Duvakina}
\affiliation{Institute of Materials, Laboratory of Nanoscale Magnetic Materials and Magnonics (LMGN), École Polytechnique Fédérale de Lausanne (EPFL), 1015 Lausanne, Switzerland}

\author{V. Karakhanyan}
\affiliation{Université Marie et Louis Pasteur, CNRS, Institut FEMTO-ST, F-25000 Besancon, France}

\author{M. Xu}
\affiliation{Institute of Materials, Laboratory of Nanoscale Magnetic Materials and Magnonics (LMGN), École Polytechnique Fédérale de Lausanne (EPFL), 1015 Lausanne, Switzerland}

\author{M. Suarez}
\affiliation{Université Marie et Louis Pasteur, CNRS, institut FEMTO-ST, F-25000 Besancon, France}

\author{A. J. M. Deenen}
\affiliation{Institute of Materials, Laboratory of Nanoscale Magnetic Materials and Magnonics (LMGN), École Polytechnique Fédérale de Lausanne (EPFL), 1015 Lausanne, Switzerland}

\author{M. Raschetti}
\affiliation{Université Marie et Louis Pasteur, CNRS, institut FEMTO-ST, F-25000 Besancon, France}

\author{A. Mucchietto}
\affiliation{Institute of Materials, Laboratory of Nanoscale Magnetic Materials and Magnonics (LMGN), École Polytechnique Fédérale de Lausanne (EPFL), 1015 Lausanne, Switzerland}

\author{T. Grosjean}
\email{thierry.grosjean@univ-fcomte.fr}
\affiliation{Université Marie et Louis Pasteur, CNRS, institut FEMTO-ST, F-25000 Besancon, France}

\author{D. Grundler}
\email{dirk.grundler@epfl.ch}
\affiliation{Institute of Materials, Laboratory of Nanoscale Magnetic Materials and Magnonics (LMGN), École Polytechnique Fédérale de Lausanne (EPFL), 1015 Lausanne, Switzerland}
\alsoaffiliation{Institute of Electrical and Micro Engineering, EPFL, 1015 Lausanne, Switzerland}

\begin{document}

\begin{abstract}

The growing demand for dense, energy-efficient, and high-frequency signal processing continues to drive device miniaturization. While downscaling remains a central challenge, magnons offer a promising solution as nanoscale signal carriers, supporting broadband operation from GHz to THz without moving charge carriers and generating Joule heating. However, their integration at the nanoscale is limited by conventional electrical excitation based on coplanar waveguides, which require metal pads few to hundreds of micrometres in size. Here, we demonstrate tunable magnon emission into a yttrium iron garnet film by focusing microwave-modulated laser light onto an integrated Au nanodisc. Using inelastic light scattering spectroscopy, we observe magnons whose frequencies match the optical modulation frequencies in the GHz frequency regime. The largest magnon amplitudes are found for circularly polarized laser light and specific nanodisc diameters consistent with a plasmon-enhanced inverse Faraday effect. These results establish plasmonic nanoantennas as reconfigurable nanoscale magnon sources, enabling broadband signal generation governed entirely by optical modulation.
\end{abstract}

\section{INTRODUCTION}
Magnonics offers a unique way to develop microwave signal processing on the nanoscale while reducing joule heating \cite{Khitun_2010, Chumak2014, Abdulqader_2020}. 
A condition for downscaling spin wave-based functionalities is to engineer tunable sources of spin waves (magnons) whose lateral extensions do not exceed a few (tens of) nanometers. A small emitter offers large wave vector components $k=2\pi/\lambda$, where $\lambda$ is the magnon wavelength \cite{Yu2013-xj}.
Scalable nanoemitters for launching short-wavelength magnons at specific microwave frequencies and different locations on a chip are urgently needed. Making use of an optical trigger would even allow for ultrahigh bandwidth, avoidance of electrical wiring and interconnects with high-speed photonic technology.  
Recent approaches to magnon emission from nanoscale objects in the few GHz frequency regime involved magnetic skyrmions \cite{Chen_Jilei_2021_Haiming}, magnetic vortices and domain walls \cite{Wintz2016,Mayr2021} existing in magnetic multilayers. The emitters required the complex engineering of magnetic textures which were then exposed to global radiofrequency magnetic fields and generated propagating spin waves in the surrounding films. 
Photonics has already enabled methodologies for stimulated magnetization reversal \cite{C9NR01628G,10.1063/5.0262327} and the generation of magnons by the use of light pulses \cite{wang:prb07,lenk:prb10,au:prl13}. A key mechanism relied on the inverse Faraday effect (IFE) produced with single or periodic light pulses in the optical regime\cite{chernov:ol17,jackl:prx22,savochkin:scirep17}. Achieving significant IFE in a transparent magnetic materials such as ferrimagnetic yttrium iron garnet (YIG) would require high-intensity light pulses \cite{Jackl_and_Belotelov_2017_PRL}, implying unintentionally large consumption of energy for producing spin waves. Being diffraction limited, the resulting laser spots do not allow for ultracompact magnon emitters. 
Recently, sub-wavelength gratings have been successfully used to locally excite spin waves beyond the diffraction limit \cite{chernov:nl20}. Here, the tightly confined light spots were obtained through coherent scattering across a large array of sub-wavelength grooves. An individual nano-optical source for selectively excited spin waves has not yet been presented.   
An alternative strategy for overcoming the diffraction limit and enhancing light–matter interaction at the nanoscale involves the use of surface plasmons, which enable tight localization of light and significant enhancement of optical intensity, respectively.\cite{10.1063/5.0262327,garcia:jpd11, novotny:np11} Recent theoretical studies have predicted that the IFE can be both confined and amplified within plasmonic nanoantennas,\cite{nadarajah:ox17, mondal:prb15, hurst:prb18, sinha:acsphot20, hamidi:oc15, karakhanyan:ol21, karakhanyan:osac21, yang:acs21, lefier:thesis, karakhanyan2022inverse, yang:nl25, gu2010plasmon, liu:nl15, dutta:ome17,singh:ox12,pineider:nl13,moocarme:nl14} promising highly localized nanoscale optomagnetic sources, or “optomagnets”\cite{karakhanyan:ol21}. Experimental demonstrations of plasmon-enhanced optomagnetism have recently emerged, particularly in gold nanoparticle colloids and gold nanodisc arrays, using both femtosecond pulse excitation\cite{Parchenko_2025, gonzalez:nanophot24, cheng:np20} and continuous-wave (CW) illumination.\cite{cheng:nl22} However, despite these advances, plasmon-enhanced optomagnetism has not yet been employed to actively generate or control magnetic excitations such as spin waves in YIG.

\begin{figure}[H]
    \centering
    \includegraphics[width=1\linewidth]{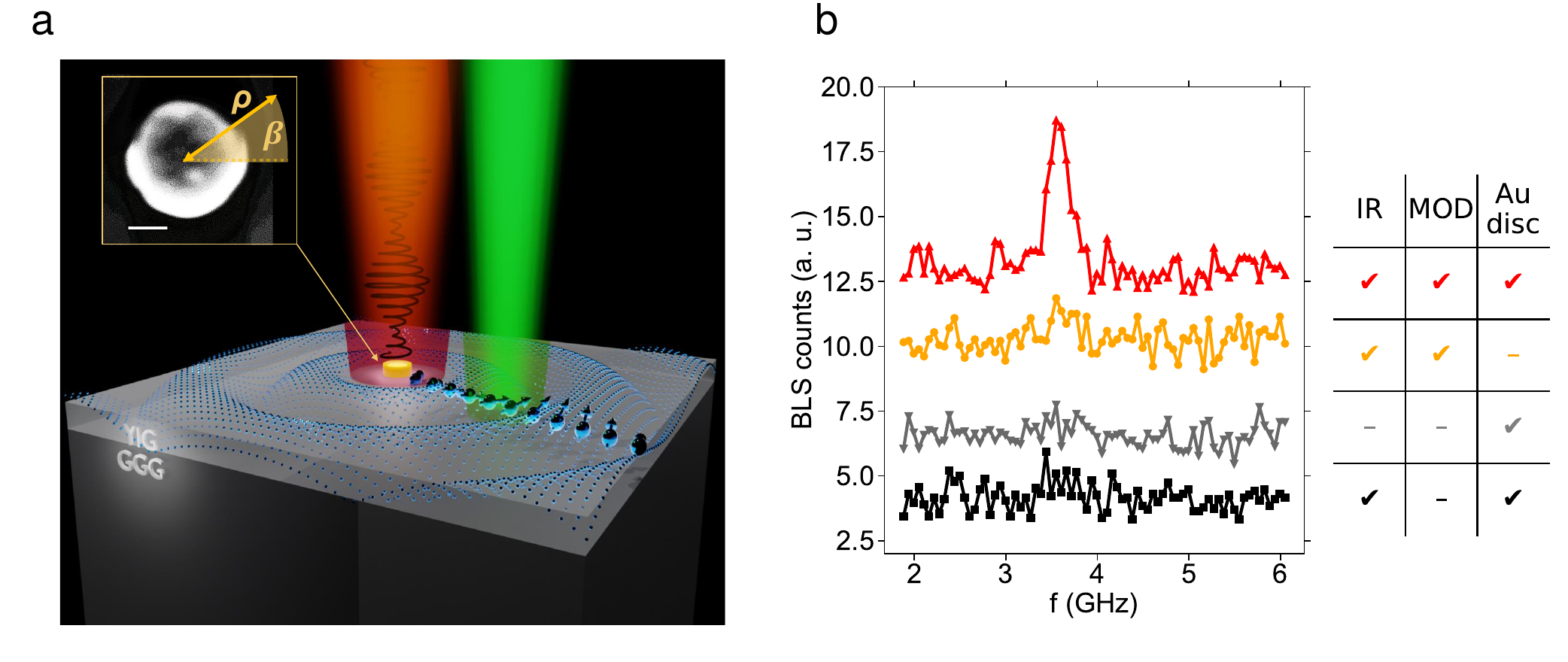}
%\end{figure}
%\begin{figure}[H]
    \includegraphics[width=1\linewidth]{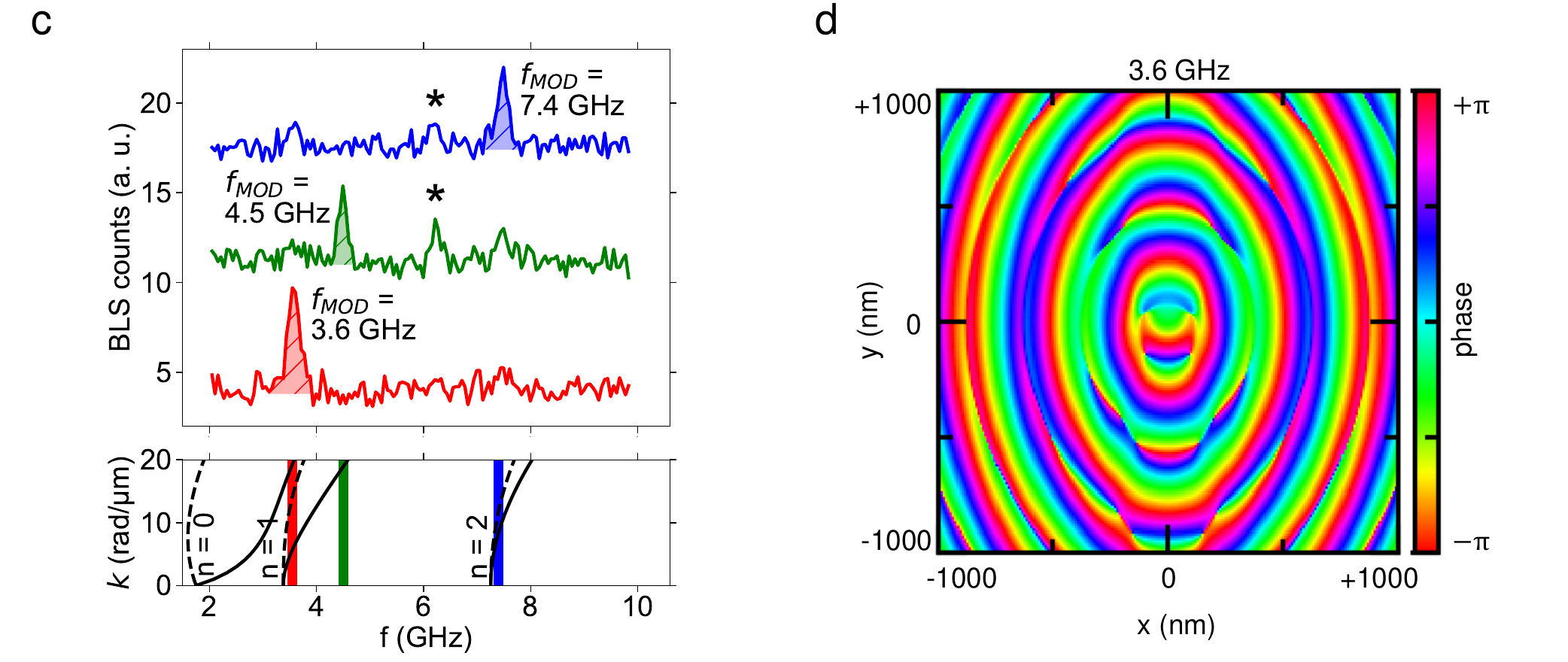}
    \captionof{figure}{\textbf{Broadband excitation of monochromatic spin waves by an optomagnetic nanoemitter.}\\
    {\textbf{a}}, Spin waves (blue spheres and wavefronts) are optomagnetically excited by infrared (IR) light (red) focused on top of the Au nanodisc (gold) and detected in the YIG film via inelastic light scattering. The probe laser (green) is positioned at a radial distance $\rho$ and angle $\beta$ in a cylindrical coordinate system centered on the disc. Inset: SEM image of a 130 nm-diameter Au nanodisc on YIG (Nanodisc I on sample I). Scale bar: 50 nm. {\textbf{b}}, BLS spectra recorded at Position 1 ($\rho = 450$ nm, $\beta = -30^\circ$) for four different conditions indicated in the table: with and without IR beam and modulation (MOD) at $f_{\text{mod}}$ = 3.6 GHz. For red, gray and black spectra the IR laser was on Nanodisc I. The orange spectrum was taken when the lasers were positioned on bare YIG without a nanodisc, maintaining $\rho = 450$ nm.  {\textbf{c}}, BLS spectra taken on the sample I (Position 1) with enhanced magnon signals (shaded peaks) at $f_{\text{mod}}$ of 3.6 GHz (red), 4.5 GHz (green), and 7.4 GHz (blue). The small peaks at 6.2 GHz (indicated with an asterisk) did not depend on $H$ and were attributed to a parasitic line of the green laser. Bottom inset: Spin-wave dispersion relations for a 113 nm-thick YIG film in Damon-Eshbach (solid line), and backward-volume (dashed line) geometry for the thickness modes $n = 0, 1, 2$. Vertical lines indicate the different $f_{\text{mod}}$ of the IR laser. {\textbf{d}}, Micromagnetic simulation of the dynamic magnetization (phase map) around a plasmonic nanodisc under CW excitation at 3.6 GHz. In panels {\textbf{b}} to {\textbf{d}} an in-plane magnetic field $\mu_0H=20~$mT existed along the $\beta = 0^\circ$ direction.}
    \label{fig:figure1}
\end{figure}

In this work, we demonstrate a nanometer-scale optical source of spin waves by coupling a microwave-modulated continuous-wave (CW) laser to an individual Au nanodisc on YIG (Figure \ref{fig:figure1}a). Under circularly polarized illumination, the nanodisc converts most efficiently the microwave-modulated optical field into monochromatic spin waves. By controlling the modulation frequency, we drive the emission of spin waves across distinct magnon bands in the gigahertz regime, demonstrating the tunable and frequency-selective excitation. Our observations indicate that at its plasmon resonance a Au nanodisc functions as a nanoscale optomagnet \cite{karakhanyan:ol21}. Additionally, consistent with prior observations,\cite{cheng:nl22} we find that the presence of an external magnetic field lifts the handedness degeneracy of the resonant IFE via static Lorentz forces, allowing for spin-wave generation even under linearly polarized light. Our approach introduces a novel method for tunable optomagnonics. In turn, spin waves provide a novel way for investigating plasmon-induced optomagnetic effects on the nanoscale.

\section{RESULTS AND DISCUSSION}
%\section{RESULTS}

\subsection{Optical source of spin waves}

Figure \ref{fig:figure1}a shows a schematic of the experiment. For illustration purposes, we sketch isotropic magnon emission from the central disc. The investigated samples consist of individual gold nanodiscs with diameters $D$ varying from 70 to 190 nm, fabricated using e-beam lithography on top of a 113-nm thick Yttrium Iron Garnet (YIG) film. Optical excitation is provided by a right circularly polarized (RCP), amplitude-modulated laser beam at a wavelength of 808 nm and modulation frequencies in the gigahertz range. 
%This wavelength and nanodisc size range are chosen to induce a localized surface plasmon resonance (LSPR) \cite{karakhanyan:ol21}. 
The design of nanodiscs allows for a localized surface plasmon resonance at (or near) the wavelength of the excitation IR laser (See Figure S1 in the Supporting Information).
To probe spin waves, we apply micro-focus Brillouin light scattering (BLS) spectroscopy using a linearly polarized 532-nm-wavelength laser. Both laser beams are focused on the sample through the same objective lens. Their incidence is normal to the surface. The excitation beam is centered around a nanodisc, while the probe beam is positioned at a distance $\rho$ from the disc center and oriented at an angle $\beta$, as illustrated in the inset of Figure~\ref{fig:figure1}a (top right). It shows a scanning electron microscope (SEM) image of a nanodisc  with $D=130~$nm (Nanodisc I). In the following the laser beam configuration with parameters $\rho = 450$ nm and $\beta = 30^\circ$ is referred to as Position 1. It serves as the primary configuration used in the main experiments (see Section S1 of the Supporting Information). All measurements are performed under an in-plane magnetic field $H$ applied along the $\beta = 0^\circ$ direction. For the recorded BLS spectra, the data counts were normalized according to the procedure described in the Methods Section, ensuring a systematic approach for direct comparison of measurements under different optomagnetic excitation conditions and across different nanodiscs. %Additional experimental details can be found in Section S1 of the Supporting Information.

Figure \ref{fig:figure1}b shows that spin wave spectra taken for $\mu_0H = 20~$mT at Position 1 reveal a BLS signal enhancement under specific conditions. The enhancement occurs in the presence of an Au disc under illumination by the RCP laser beam modulated at a frequency of $f_{\text{mod}}$ = 3.6 GHz. This frequency lies within the range of the allowed magnon band in YIG at $\mu_0H= 20~$ mT, as determined by the Kalinikos-Slavin dispersion relation \cite{Kalinikos_1986} (bottom panel of Figure \ref{fig:figure1}c). The BLS peak enhancement was absent when spectra were taken (i) on a bare YIG film while keeping all other parameters unchanged, (ii) without IR illumination, or (iii) without modulation of the IR beam. Additionally, a signal enhancement is absent at $\mu_0H = 100~$ mT (see Supporting Information, Figure S3). At this field, the bottom of the magnon band is located above the IR laser beam modulation frequency.

To investigate frequency control of magnon excitation, we took BLS spectra for which the modulation frequency was tuned. The BLS spectra were measured at Position 1 on Nanodisc I for $\mu_0H = 20~$ mT. In Figure \ref{fig:figure1}c we observe enhanced BLS signals at frequencies matching the modulation frequencies of the laser beam $f_{\text{mod}}$: 3.6 GHz (red), 4.5 GHz (green), and 7.4 GHz (blue), each highlighted with shading. Here, we specifically targeted frequencies that intersect allowed magnon bands. We depict spin-wave dispersion relations for a 113 nm-thick YIG film in the Damon-Eshbach (solid line) and backward-volume (dashed line) geometries and for Perpendicular standing spin-wave (PSSW) modes with $n = 0, 1$, and $2$ in the bottom inset of Figure~\ref{fig:figure1}d \cite{Kalinikos_1986}. Colored lines indicate the modulation frequency of the IR light. The displayed regime of wave vectors $k$ is accessible by microfocus-BLS \cite{PhysRevResearch.2.033427}. A field-independent peak at 6.2 GHz (marked with an asterisk) is attributed to a spurious laser-induced signal. 

To further interpret our data, we performed micromagnetic simulations of the YIG film under continuous excitation  (Figure \ref{fig:figure1}d) \cite{vansteenkiste_design_2014}. In the simulations, we considered a 113-nm-thick YIG film with 9 cells along the thickness in order to model the perpendicular standing spin waves (PSSWs). A constant bias field of $\mu_0H= 20~$mT was applied along the $x$-direction, and absorbing boundaries where the damping is quadratically increased near the system boundaries minimized back reflections. To consider the dynamic field underneath the Au disc, we computed the spatial distribution of the optomagnetic field $h_0(r)$ produced by a plasmonic disc (see Supporting Information Figure S4). This profile was normalized to have a maximum out-of-plane component strength of 6.3 mT. The full time-dependent optomagnetic field of the disc $h_{\rm plasmon}$ was constructed by multiplying the static profile with a dependent function leading to $h_{\rm plasmon}(r,t)=h_{0}(r)|sin(2\pi(f/2)t|$, with $f=3.6$~GHz and $t$ the time. Starting from the ground state in the absence of any excitation, we introduced the localized dynamic field $h_{\rm plasmon}(r,t)$ and ran the simulation until the steady state was reached in YIG. Additional details on the simulations are provided in Methods Section. Figure \ref{fig:figure1}d shows the distribution of the phase of the dynamic magnetization $\delta m_{\rm z}$ of the second top-most layer in YIG. The spatial distribution of the dynamic magnetization and the norm of its Fourier transform are shown in Figures S5 and S6 of the Supporting Information. The simulations show a regular pattern of coherent magnons in all directions. They suggest that magnons of different wavelengths $\lambda$ are coherently excited at 3.6 GHz and propagate differently in in-plane directions. The wavelengths $\lambda=2\pi/k$ are consistent with the branches obtained by the Kalinikos-Slavin formalism at 3.6 GHz. The spatial distribution will be discussed in detail later.

%\subsection{BLS counts as a function of the nanodisc diameter}
\subsection{Tuning the plasmon resonance of nanodiscs}
\begin{figure}[H]
    \centering
    \includegraphics[width=\linewidth]{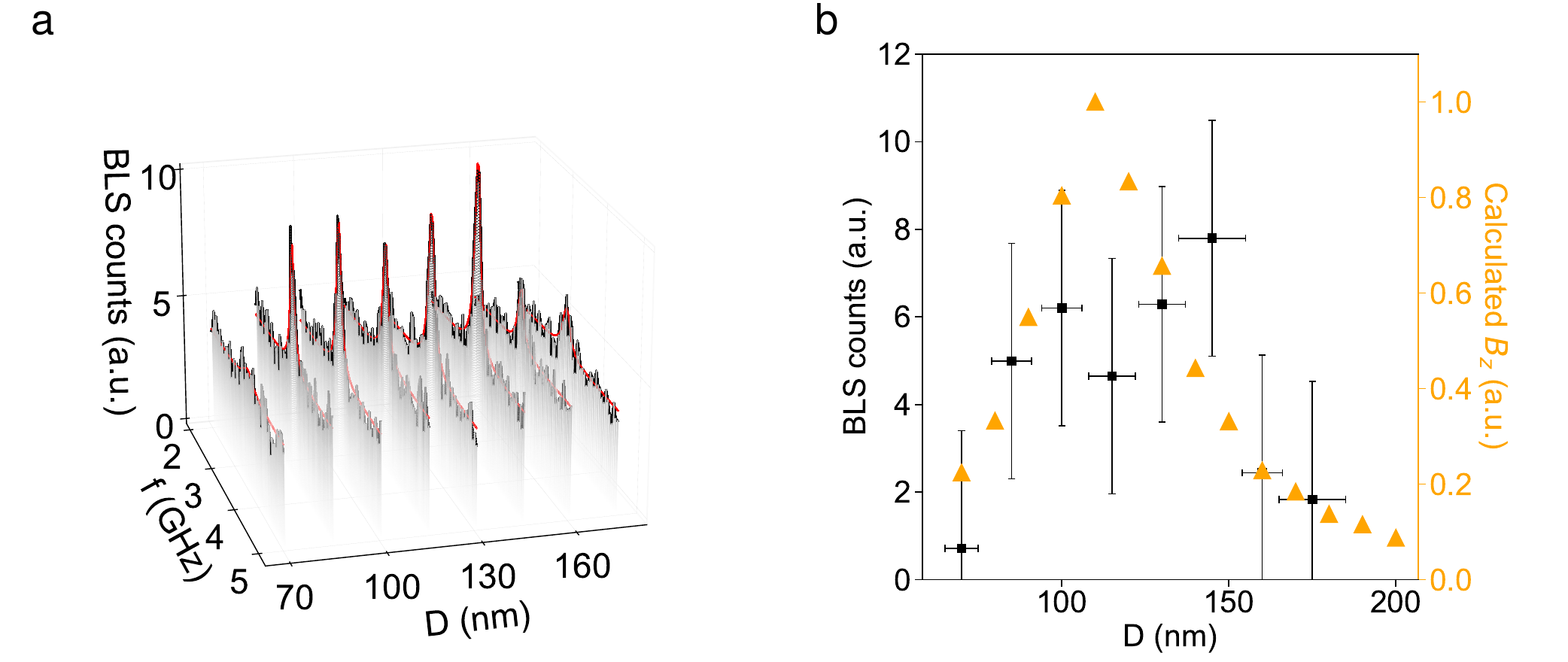}
    \captionof{figure}{
    %\textbf{Comparison of BLS signals and simulated magnetic field amplitude generated by optomagnet as a function of Au disc diameter $D$.}\\
    \textbf{Variation of the Nanodisc Diameters and Plasmon Resonances.}\\
    %{\textbf{a}}, BLS counts measured on gold discs with nominal diameters $D$ ranging from 70 to 190 nm, under a 20 mT in-plane magnetic field and 3.6 GHz modulation, at Position 1. The red curves represent Lorentzian fits to the spectra.
    {\textbf{a}}, Brillouin light scattering (BLS) spectra measured for eight Au nanodiscs with diameters ranging from 70 nm to 190 nm, under 20 mT in-plane magnetic field and modulation at $f_{\text{mod}} = 3.6~$GHz of the IR beam (measured at Positions 1). The red curves represent Lorentzian fits to the spectra.    
    %{\textbf{b}}, Amplitudes extracted from the Lorentzian fits in panel {\textbf{a}} are plotted as black squares. The yellow data points represent the values of the optomagnetic field calculated from the theoretical model for comparison.
    {\textbf{b}}, Black squares: amplitudes of BLS intensity extracted from the Lorentzian fits of each spectrum in panel {\textbf{a}} as a function of nanodisc diameter $D$. Horizontal and vertical error bars represent fabrication uncertainties in nanodisc diameter and the standard deviation of the spectral maxima, respectively. Orange triangles: calculated amplitudes of the normalized optomagnetic field component $B_z$ parallel to the nanodisc axis (See Methods Section).
    }
    \label{fig:figure2}
\end{figure}

Figure \ref{fig:figure2}a presents normalized BLS spectra measured 
%on gold discs with nominal diameters (D) 
for eight different individual nanodisc antennas of increasing diameter ($D$)
ranging from 70 to 190 nm under an in-plane magnetic field of 20 mT. 
The relative positions $\rho$ of the excitation and probe laser beam 
%remains consistent with that in Position 1, and the modulation frequency is set to $f_{\text{mod}}$ = 3.6 GHz. The red curves of the Figure \ref{fig:figure2}a represent Lorentzian fits to the experimental data.  
spots are maintained as in Position 1, with the modulation frequency fixed at 3.6 GHz. The red curves of the Figure \ref{fig:figure2}a represent Lorentzian fits to the normalized experimental data. 
%To normalize the maximum BLS counts for each spectrum, the noise level—determined from the counts at D = 70 nm — is first subtracted. 
The black symbols (circles) in Figure \ref{fig:figure2}b display the amplitudes of the fitted peaks, extracted from each spectrum from \ref{fig:figure2}a as a function of the nanoantenna diameter. The calculated amplitudes of the optomagnetic field component parallel to the symmetry axis ($B_z$) of the plasmonic nanodisc are shown as yellow triangles. The diameter-dependence of this field was estimated from a simplified hydrodynamic model of the free-electron gas in gold, evaluated at the YIG/gold interface near the edge of the nanodisc (see Methods Section). \cite{karakhanyan:ol21, karakhanyan:osac21} $B_z$ exhibits a maximum at a Au disc diameter of $D = 110$~nm where the plasmon resonance occurs at a wavelength of 808~nm. The magnetic field $B_z$ enters the equation of motion of spins in underlying YIG \cite{gurevich1996magnetization} and exerts a torque to induce their precession.
The maximum magnon signal in the BLS experiments is found at a slightly higher value of $D=150~$nm.

%\subsection{Measured spatial dependence of BLS signal for circularly and linearly polarized light}
\subsection{Measured spin wave emission pattern for circularly and linearly polarized light}
\begin{figure}[H]
     \centering
     \includegraphics[width=\linewidth]{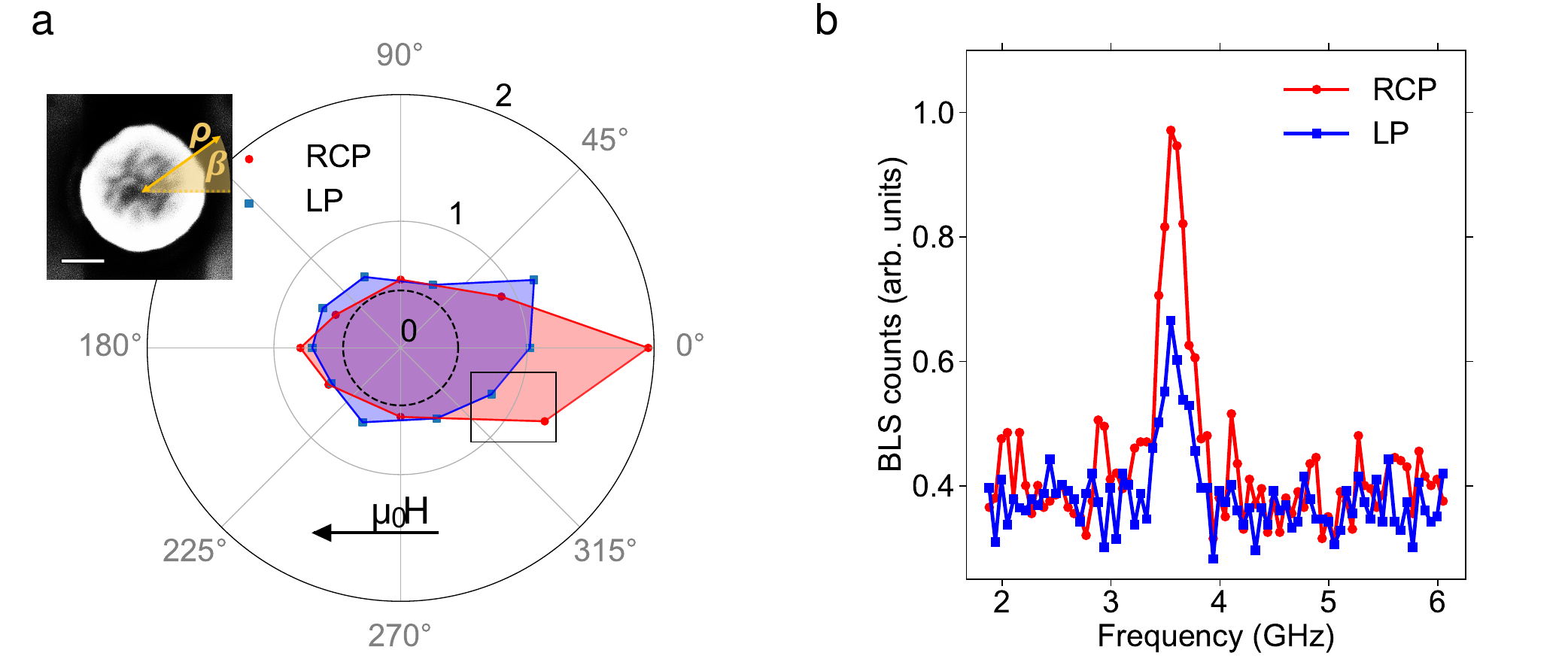}
   % \captionof{figure}{\textbf{Measured spatial dependence of BLS signal for circularly and linearly polarized IR laser excitation.}}
    \captionof{figure}{\textbf{Directional spin-wave emission from an optomagnet for both RCP and LP excitation.}\\
   %{\textbf{a}}, Polar diagram of the position dependence of maximum BLS counts in the presence of a IR laser with intensity modulated at a frequency of 3.6 GHz in an applied magnetic field of 20 mT: red - CP, blue - LP. Black dashed circle represents the BLS counts on the Position 1 without  IR laser. Inset: SEM image of the Au disc with the diameter of 130 nm (Sample II). The angle $\beta$ is indicated. Scale bar: 50 nm. {\textbf{b}}, BLS spectra taken on the Position 1 at $\beta = -30^\circ$ (black box in {\textbf{a}).}
  {\textbf{a}}, Measured spin wave emission pattern under optical excitation modulated at 3.6 GHz in the presence of a 20 mT in-plane external magnetic field. The IR excitation light is either RCP (red dots) or LP (blue dots). Black dashed circle: background BLS signal level on Position 1, in the absence of IR light. Inset: SEM image of the tested gold nanodisc with the diameter of 130 nm (Nanodisc II, which is a nominally identical Au nanodisc to Nanodisc I). The angle $\beta$ is indicated on the image. Scale bar: 50 nm. {\textbf{b}}, BLS spectra recorded for RCP and LP IR excitations, at  $\beta = -30^\circ$ (see the position marked by the black box in {\textbf{a}}).}
    \label{fig:figure3}
\end{figure}

%In Figure \ref{fig:figure1}a, we assumed isotropic magnon emission from the optomagnet; in contrast, experimental results reveal a directional emission pattern. Under CP excitation, spin-wave  emission is maximized along the $\beta = 0^\circ$ direction, which is aligned with the external magnetic field. For LP excitation, a qualitatively similar anisotropic emission is observed, albeit with an reduced overall signal amplitude. This behaviour is illustrated in Figure \ref{fig:figure3}b, which shows the spectra acquired at the directional angle $\beta = - 30^\circ$ (corresponding to the position within the black frame in Figure \ref{fig:figure3}a). The inset of the Figure \ref{fig:figure3}a shows the SEM image of a gold nanodisc (Nanodisc II), which is nominally identical to Nanodisc I. The disc exhibits unintentional edge roughness, which breaks its circular symmetry and may influence the spin-wave emission pattern. 
Position-dependent BLS spectra were recorded around a plasmonic nanodisc while maintaining the excitation light spot consistently centered to the nanoantenna's symmetry axis. A constant distance of $\rho = 450~$nm was used between the fixed IR beam and the scanning BLS probe spot. 
%The excitation beam was positioned directly on top of the antenna, while the probe beam was placed at different angles $\beta$, relative to the horizontal axis. 
The probe spot was sequentially positioned at different angles $\beta$, and a BLS spectrum was recorded at each position. This procedure was performed for both RCP and linearly polarized (LP) IR light. 
%Figure \ref{fig:figure3}a displays a polar plot showing the maximum BLS signal as a function of the angle $\beta$ for both a circularly polarized (CP) and linearly polarized (LP) IR beam. 
Figure \ref{fig:figure3}a displays a polar plot showing the amplitude of the BLS spectra recorded at each $\beta$ angle for both RCP and LP IR light. In the polar plot, the distance of a data point from the center represents the amplitude of the normalized BLS spectra, obtained by the procedure described in Methods Section. The black dashed circle represents the background BLS signal when the IR laser was switched off.
The experimental results reveal a directional emission pattern. Under circularly polarized excitation, spin-wave  emission is maximized along the $\beta = 0^\circ$ direction, which is aligned with the external magnetic field. For LP excitation, a qualitatively similar anisotropic emission is observed, albeit with a reduced overall signal amplitude. The different signal strengths are shown in the spectra of Figure \ref{fig:figure3}b acquired at $\beta = - 30^\circ$ (corresponding to the position within the black frame in Figure \ref{fig:figure3}a). The black dashed line represents the background BLS signal when the IR laser was switched off. 
The measured spin-wave emission pattern is anisotropic in contrast to a radially symmetric emission process assumed in Figure~\ref{fig:figure1}a for an isotropic medium. At 3.6 GHz different magnon modes exist, with wavelengths $\lambda$ which differ in different in-plane directions of the YIG magnetized in $x$-direction. The color-coded map shown in Figure~\ref{fig:figure1}d illustrates the anisotropic characteristics of magnon phases expected in YIG. 

%$A_\mathrm{FeGd}=3.75~\mathrm{pJ/m}$, gryomagnetic ratio $|\gamma|=176~\mathrm{rad~GHz~T^{-1}}$ and damping $\alpha=9\cdot10^{-4}$. The system was discretized into $1200\times1200\times9$ cells of dimensions $5\times5\times6.3~\mathrm{nm}^3$. To prevent spin wave reflection, the damping was icnreased quadratically to 1 on the sides.\\

%\section{DISCUSSION}
\label{sec:discussion}
Our results demonstrate frequency-selective emission of spin waves when illuminating a Au nanodisc by an amplitude-modulated IR laser which is tuned to the plasmonic resonance. By focusing a GHz-modulated circularly polarized IR light onto individual nanodiscs with a diameter of about 130~nm, we observe enhanced BLS signals at user-defined frequencies, which are consistent with allowed magnon bands in YIG. Our approach enables one to target a selected GHz frequency by tuning the IR laser modulation frequency. This feature is in contrast to ultrafast optical pumping, which excites a continuum of modes \cite{satoh:natphot12, Yoshimine2014, parchenko:apl13}. While the current modulation frequency is limited to below 10 GHz by the currently used electro-optical modulator, the technique offers wide frequency tunability with advanced modulators. Note that electro-optic modulator with bandwidths of several tens of GHz are commercially available. Together with the high spatial localization (defined by the nanodisc footprint), they %\mr{~tens of nm?, better give a number}), -> it can go for tens (we refer to one of the work) or hundreds (in our work), so I am not sure, do not want to define it here.
%\mr{~GHz?}, -> last paragrapf's sentence.
make the demonstrated optomagnetic emitters a promising platform for signal generation in integrated and reconfigurable magnonic devices. \\
%To support our experimental findings, we performed electromagnetic and micromagnetic simulations of the optical excitation of the plasmonic disc and spin-wave emission, respectively. Electromagnetic simulations of the Au nanodisc on YIG confirm strong confinement of the CP infrared field in its near-field region, with an enhanced circularly polarized E-field component penetrating into the YIG film. Using an established magneto-optical formalism for the inverse Faraday effect (IFE)\cite{karakhanyan2022inverse}, we computed the optomagnetic field $H_{\text{IFE YIG}}$ distribution in the YIG. We incorporated the spatial distribution of $H_{\text{IFE YIG}}$ profile into micromagnetic simulations. The amplitude of $H_{\text{IFE}}$ was modulated at the given GHz frequency. These simulations show that the localized oscillating magnetic field excites coherent spin waves in the YIG film. 
Our micromagnetic simulations (compare Figure ~\ref{fig:figure1}d) suggest the localized excitation of coherent spin waves belonging to the magnon bands at the selected (modulation) frequency. However, the anisotropic spatial profile of magnon amplitudes found around the nanodisc in simulations (Figure S5 in Supporting Information) is rotated by $-90^\circ$ compared to the results shown in Figure \ref{fig:figure3}a. The inset of Figure \ref{fig:figure3}a shows the SEM image of Nanodisc II which was studied. The nanodisc exhibits unintentional edge roughness, which breaks its circular symmetry and may lead to enhanced irregularities in the emitted spin-wave amplitudes around the disc.  But the origin of the remaining discrepancy  may stem from further factors like local alterations of the magnetic state of the YIG film, potentially caused by either the optomagnetic field itself \cite{Parchenko_2025}, inhomogeneous heating from the gold nanodisc \cite{Kuznetsov2024-sc,Parchenko_2025}, or the injection of spin-polarized electrons from gold into YIG.\cite{Parchenko_2025} It is possible that a combination of these effects contributes to the observed anisotropy.

Recent studies have measured anomalously large magneto-optical responses or magnetic circular dichroism in resonant aqueous colloidal particles \cite{singh:ox12,pineider:nl13,moocarme:nl14} or gold nanodiscs \cite{cheng:nl22} exposed to an external static magnetic field, due to resonant IFE. These phenomena have been observed even at low optical intensities and weak static external magnetic, on the order of 1 W.cm$^{-2}$ and 1 mT, respectively \cite{singh:ox12,moocarme:nl14}. Importantly, the presence of an external magnetic has been shown to lift the intrinsic handedness degeneracy of the IFE within the nanoantennas \cite{singh:ox12,pineider:nl13}. This property suggests the potential for an optomagnetic response even under a LP excitation. In these prior studies, where the magnetic field was aligned with the spin angular momentum of the incident photons, the observed magneto-optic effects were successfully explained by a model based on radially distributed Lorentz forces. These forces acted on the azimuthal optoinduced currents within the nanoantennas, leading to a shrinkage or expansion of the current loops depending on the handedness of the circular polarization in the presence of a static external magnetic. This mechanism provided a clear physical explanation for the lifting of handedness degeneracy of the IFE in the plasmonic nanostructures. In our case, where the static external magnetic field is orthogonal to the optical spin angular momentum, no simple analytical model can readily describe the influence of Lorentz forces on the inverse Faraday effect. A fully numerical approach, incorporating the spatial and temporal discretization of the hydrodynamic model of the free-electron gas \cite{karakhanyan:ol21,karakhanyan2022inverse,karakhanyan:osac21} and explicitly accounting for the Lorentz force induced by the static external magnetic field, would be necessary to accurately capture the underlying dynamics. Furthermore, contributions from bound electrons within the metallic lattice should also ideally be considered, as demonstrated in similar models developed to predict second-harmonic generation in plasmonic nanoantennas.\cite{scalora:pra10,ciraci:prb12}. In our fabricated nanodiscs, imperfections such as the edge roughness may lead to deviations of surface currents from ideal circular loops, introducing additional complexity to the underlying optomagnetic processes.

The local minimum observed in the response of our optomagnetic system at $D=110~$nm in Figure~\ref{fig:figure2}b is in contrast to the peak predicted by our optomagnetic field calculations. A local minimum was reported in a previous study of plasmon-induced optomagnetism in arrays of nanodiscs on a thin magnetic film.\cite{Parchenko_2025} This minimum was attributed to local modifications of the magnetic state of the magnetic film, resulting from a combination of photothermal effects and optomagnetic fields generated by the plasmonic nanoantennas. Unlike the study by Parchenko et al.,\cite{Parchenko_2025} which employed femtosecond laser pulse excitation, our experiments use CW optical excitation, which, a priori, can induce temperature-induced effects. Consistent with arguments in Ref. \cite{Parchenko_2025}, we observe that our observed spin-wave excitation efficiency reaches its maximum on the sides of the plasmon resonance, where light absorption by the resonant surface plasmon remains relatively modest. Here, the spin-wave excitation efficiency would arise from a trade-off between the plasmon-enhanced IFE and a possible alteration of the YIG magnetic state. It is important to note that the frequency of the PSSW peak in YIG measured very close to the nanoantenna remained unchanged with and without IR excitation, indicating that the saturation magnetization and spin-wave dispersion relation in YIG was not modified in our experiments (see Supporting Information Figure S7). This is different from studies performed by Kuznetsov et al.\cite{Kuznetsov2024-sc}, who illuminated extended nanodisc arrays spanning several tens of micrometers. Using large laser power, they optimized the collective thermoplasmonic effect for modifying spin-wave transport in YIG. We do not expect such an effect in the case of our individual nanodiscs.

Building on our findings, several research directions emerge. On the one hand, because the excitation frequency is set by an external modulator, multiple nanodisc emitters could be driven at different frequencies simultaneously by using multi-tone
%\mr{multi-tone? multicolor?}
modulation signals. This could enable frequency-division multiplexing in magnonic networks, where different channels of information are carried by spin waves of different frequencies through the same medium \cite{Vogt2014}. Moreover, by phasing the driving signals appropriately, one could create phased arrays of spin-wave nanoemitters to direct spin-wave beams and achieve interference-based magnonic logic functions \cite{Papp2021}. The small size of the emitters means they can be placed close enough to allow for nanoengineered interference patterns. The optically induced magnetic effects from separate emitters can induce propagating spin waves and, in the nonlinear regime, influence the characteristics of the already propagating ones ("gating"). On the other hand, our experiments show that spin waves offer a platform for probing optomagnetic phenomena at the nanoscale.  We demonstrate their probing capability by showing their unique sensitivity to the resonant IFE in a single nanoantenna. In the context of plasmon-induced optomagnetism, spin waves can act as dynamic sensors that reveal how localized optical fields modify magnetic order. This dynamic probing capability makes spin waves suited for exploring ultrafast and spatially confined optomagnetic interactions, providing a pathway to a deeper understanding and potential control of light-matter interactions in hybrid photonic-magnetic systems.
%In summary, 
\section{CONCLUSIONS}

We demonstrate infrared-to-GHz magnon coupling via optical modulation in a metal disc, with all dimensions confined to the nanometer scale. The plasmon-enhanced, IFE-driven mechanism bridges photonics and magnonics at the nanoscale, enabling light-based control of spin information and opening a path toward low-power, all-optical magnonic technologies for scalable GHz–THz information processing. Electrical wiring or integrated microwave antennas are not required. Instead of pulsed perturbation, our approach used amplitude-modulated CW IR light to periodically drive spin precession, resulting in localized magnon emission at a defined frequency. Au discs with differently engineered plasmon resonances and wide-band electro-optical modulators promise multi-frequency operation and multiplexing on the nanoscale in multi-emitter circuit layouts. Notably, our technique operates with a standard continuous-wave laser of 3 mW power, combined with an electro-optical modulator — eliminating the need for complex, high-peak-power laser systems used previously.

\section{METHODS}

\subsection*{Sample fabrication}

The samples were prepared on a 113 nm-thick yttrium iron garnet (YIG) film grown by liquid phase epitaxy on polished gadolinium gallium garnet (GGG) (111) substrates (Matesy GmbH, Jena, Germany). Plasmonic nanodiscs were defined on the insulating YIG surface using electron beam lithography (EBL). A polymethyl methacrylate (PMMA) resist was spin-coated and patterned by EBL. After development, a 5 nm titanium (Ti) adhesion layer and a 50 nm gold (Au) layer were deposited via electron beam evaporation. Nanodiscs were subsequently formed by a standard lift-off process in hot N-Methyl-2-pyrrolidone (NMP). The quality of the pattern transfer and the geometry of the fabricated structures were characterized by scanning electron microscopy (SEM). The diameters of the plasmonic nanodiscs ranged from 70 nm to 175 nm. The nanodisc geometry is engineered to match the LSPR with the excitation wavelength to enhance field confinement.

\subsection*{Optical bench}

\begin{figure}[H]
    \centering
    \includegraphics[width=\linewidth]{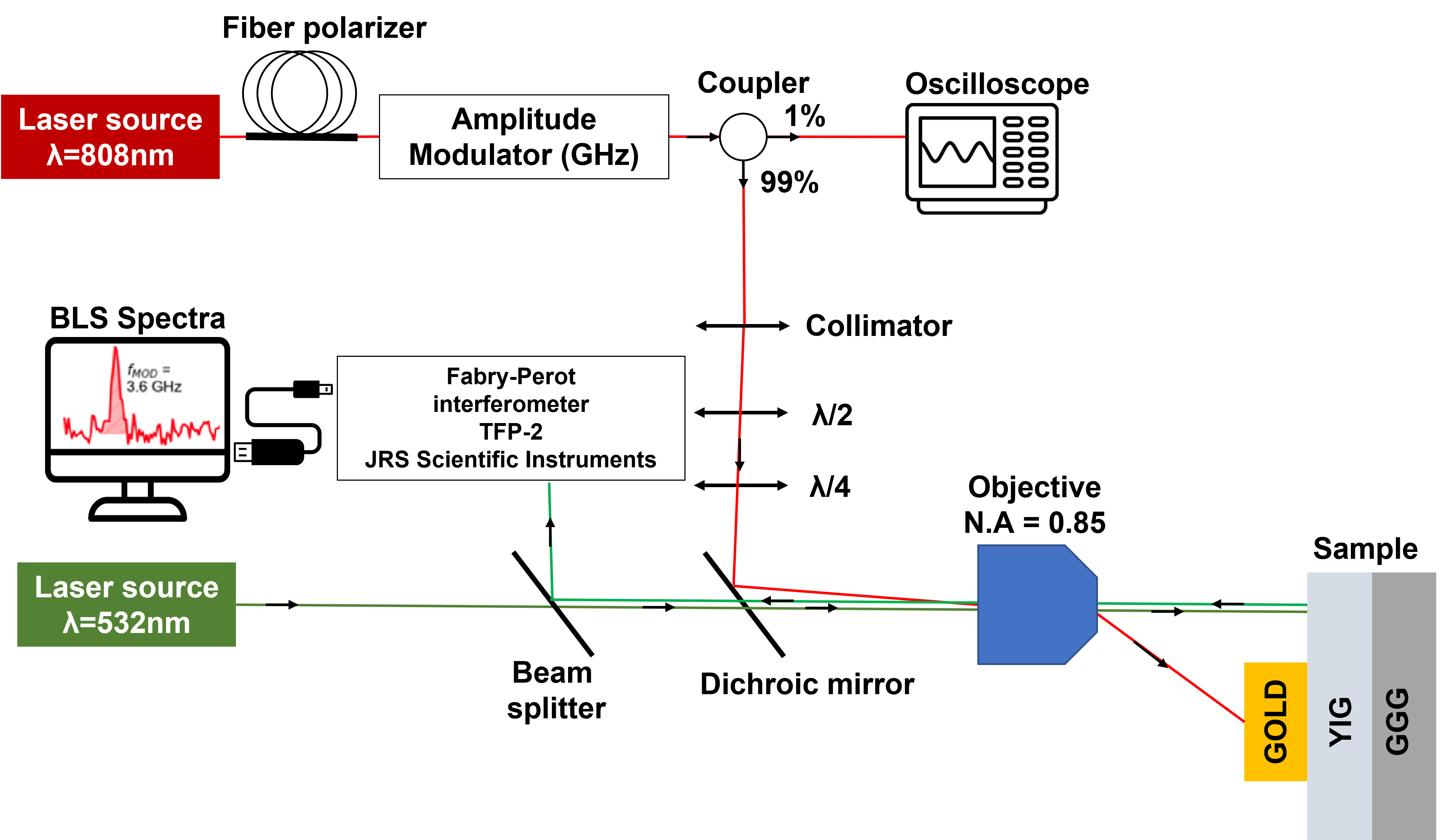}
    \caption{Experimental setup}
    \label{fig:exp_setup}
\end{figure}
The laser source (LP808-SA60) emits an 808 nm wavelength laser beam, which passes through a fiber polarizer (ILP780PM-APC), then is directed toward a GHz Modulator (EXAIL: NIR-MX800-LN-10), and finally to the coupler (Thorlabs). 
At the coupler, part of the beam (1\%) is directed to the oscilloscope to monitor the modulation process, while the other portion (99\%) is sent to a collimator. 
After the collimator, the beam passes through $\lambda/2$ and $\lambda/4$ (Thorlabs) waveplates adjusted to achieve right circular polarization. 
Then, a RCP beam reflected from the dichroic mirror (DMSP650) travels to an objective (N.A = 0.85, LCPLFLN100xLCD), which focuses the beam onto the gold nanodisc.
The nanodisc's absorption of the laser light induces a dynamic (modulated) magnetic field in the underlying YIG layer via the Inverse-Faraday effect.

\subsection*{Brillouin light scattering spectroscopy}

The spatial distribution of magnon signals was measured using a scanning micro-focus Brillouin light scattering spectroscopy (µ-BLS) setup at room temperature. A monochromatic continuous-wave solid-state laser with a wavelength of 532 nm and power of 1 mW was focused with a lens having a numerical aperture of 0.85 onto a diffraction-limited spot. Backscattered light was analysed using a six-pass Fabry–Pérot interferometer TFP-2 (JRS Scientific Instruments). The BLS signal was assumed to be proportional to the square of the amplitude of the dynamic component of the out-of-plane magnetization at the position of the laser spot. The sample was mounted on a closed-loop piezo-stage. The positioning system was stabilized by a custom-designed active feedback, providing long-term spatial stability. Permanent magnets were used to apply external magnetic fields in the plane of the YIG sample.
To ensure consistent comparison across measurements, the recorded BLS spectra were normalized following a systematic procedure. Specifically, the raw BLS counts were divided by the number of counts in the elastic (zero-frequency) peak for each measurement, compensating for variations in laser intensity. The resulting values were then multiplied by 100,000 to scale the normalized data to yield consistent, readable arbitrary units.

\subsection*{Optomagnetism}

\begin{figure}[H]
     \centering
     \includegraphics[width=\linewidth]{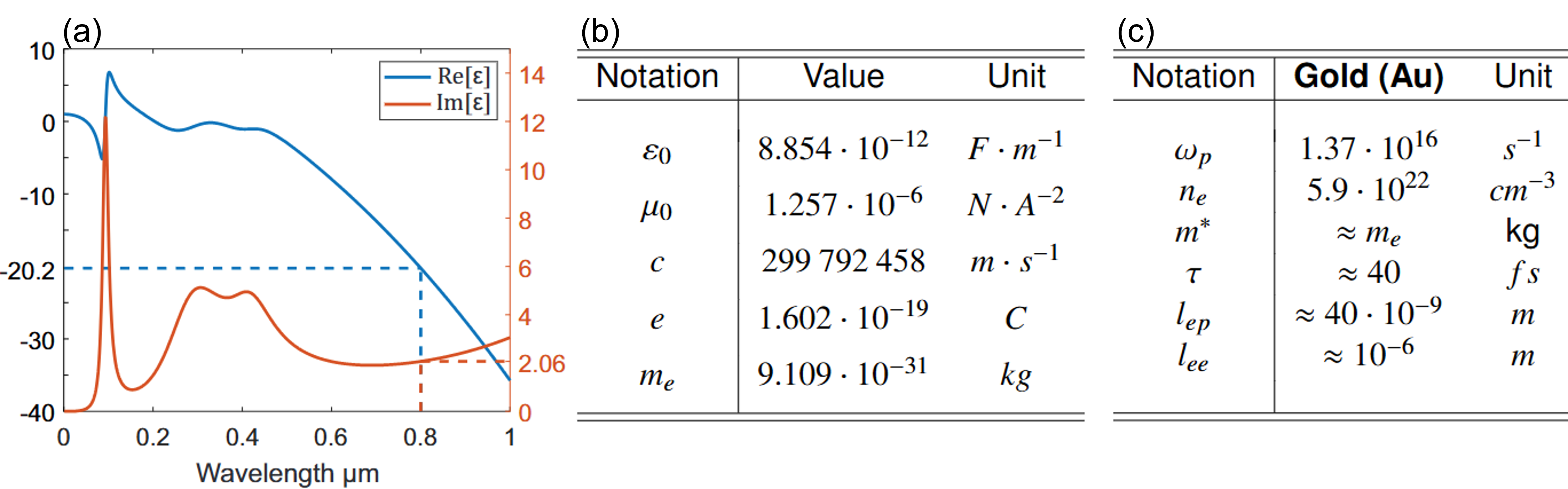}
    \caption{    \textbf{(a)} Wavelength dependent permittivity of gold, 
    \textbf{(b)} physical constants, where $m_e$ is the electron mass, $\varepsilon_0$ is the vacuum permittivity, $\mu_0$ is the vacuum permeability, $c$ is the speed of light in vacuum, $e$ is the elementary charge. 
    \textbf{(c)} Optical constants of gold. $\omega_p$ represents the plasma frequency, $n_e$ is the free electron concentration, $m^*$ is the effective mass of electrons, $\tau$ is the mean collision time, $l_{ep}$ is the mean collision length, $l_{ee}$ is the electron-electron collision length.}
    \label{fig:gold_prop}
\end{figure}

The simulation of the optomagnetic field was performed using a self-developed model, as described in Ref\cite{karakhanyan:ol21,karakhanyan2022inverse}. This model is designed to predict the nonlinear response of metals under optical illumination. Assuming that nonlinear processes in metals are primarily governed by free electrons, we employ a hydrodynamic description to model their behavior. The optical properties and physical constants of gold used in the simulations are presented in Figure \ref{fig:gold_prop}.

The computational approach consists of three sequential steps. First, the optical field distribution within the nanodisc is calculated, typically using the Finite-Difference Time-Domain (FDTD) method. Second, based on this optical field distribution, the induced currents in the nanodisc are determined. Finally, in the third step, the optomagnetic field is computed using the Biot-Savart-Laplace law. 

\subsection*{Micromagnetic simulations of magnetization dynamics}

Micromagnetic simulations were performed using the Mumax3 code \cite{vansteenkiste_design_2014} which solves the Landau-Lifshitz-Gilbert (LLG) equation following the finite difference method. For YIG, the magnetic parameters were set as saturation magnetization $M_\mathrm{S}=140~\mathrm{kA/m}$, exchange stiffness $A_\mathrm{exc}=3.75~\mathrm{pJ/m}$, gryomagnetic ratio $|\gamma|=176~\mathrm{rad~GHz~T^{-1}}$ and damping $\alpha=9\cdot10^{-4}$. The system was discretized into $1200\times1200\times9$ cells of dimensions $10\times10\times12.6~\mathrm{nm}^3$.\\
\indent
The equilibrium magnetization was found according to the following protocol. First, the magnetization of the system was initialized to a random configuration. Then, a large in plane field of 1 T was applied along the positive x-direction. This field was tilted out-of-plane by 1 degree. Then, the field was incrementally swept to -1 T and back to 1 T. In between increments, the system was equilibrated in two steps: first by minimizing the energy using the conjugate gradient method, and subsequently by solving the LLG equation without precessional term. The resulting equilibrium states were used as input for dynamic simulations.\\
\indent
To simulate the generation of magnons by coupling to the plasmonic nanoantenna, we used the spatial distribution of the dynamic magnetic field inside the YIG produced by the nanoantenna when placed on top of the YIG film
%following the procedure in S2.1. 
This magnetic field profile was multiplied with a time-dependent factor $|\sin(2\pi (f/2)t)|$ with $f$ the frequency of the time dependent field. The time-evolution of the excited magnons was found by solving the LLG for the YIG system using this dynamic magnetic field until the steady state was reached. Absorbing boundaries where the damping increases quadratically to 1 were used at all boundaries to prevent spin wave reflections.\\ 
\indent 
For comparison with the experiment, we focused on the configuration with applied static field of $20$ mT along the positive x-direction and an excitation frequency of $f=3.6$ GHz. The simulation was run for a total time of 41.6 ns. A snapshot in time of the steady state is shown in Figure S5 of the Supporting Information. To extract the phase component of the magnons at 3.6 GHz, we performed a Fast Fourier transform (FFT) over the time coordinate $t$ of the dynamic magnetization component $\delta m_{\rm z}(t, x, y, z)$, defined as $\delta m_{\rm z}(t, x, y, z)=m_{\rm z}(t, x, y, z)-m_{\rm z}(0, x, y, z)$, with $m_z(t, x, y, z)$ the z-component of the magnetization at time $t$ and at coordinates $x, y, z$. The phase shown in Figure \ref{fig:figure1}d is given by the argument of the complex magnetization $\tilde{m}_{\rm z}(f, x, y, z)$ evaluated at frequency $f=3.6$ GHz. Similarly, the wavevector distribution of the excited magnons was analyzed by performing a two dimensional FFT over the in-plane spatial coordinates $x,y$ of the dynamic magnetization $\delta m_{\rm z}(t, x, y, z)$. The resulting norm of the complex magnetization (Figure S6 of the Supporting Information) reveals the presence of two modes, which is consistent with the Kalinikos-Slavin dispersion.

\begin{acknowledgement}

The authors gratefully acknowledge Ulrich Fischer and Alessandro Ciraci for insightful discussions.
This work was supported by the SNSF via grant number 197360, the French Agence Nationale de la Recherche under projects  NANOPTiX (ANR-18-CE42-0016) and CIFOM (ANR-23-CE42-0021), the French-Swiss Collegium SMYLE, the EQUIPEX+ SMARTLIGHT platform (ANR-21-ESRE-0040), the EQUIPEX+ NANOFUTUR (ANR-21-ESRE-0012) and the EIPHI Graduate School (ANR-17-EURE-0002). This work was also supported by the French Renatech network, MIMENTO technological facility, and the Région Bourgogne Franche-Comté.

\end{acknowledgement}


%\bibliography{achemso-demo}

\providecommand{\latin}[1]{#1}
\makeatletter
\providecommand{\doi}
  {\begingroup\let\do\@makeother\dospecials
  \catcode`\{=1 \catcode`\}=2 \doi@aux}
\providecommand{\doi@aux}[1]{\endgroup\texttt{#1}}
\makeatother
\providecommand*\mcitethebibliography{\thebibliography}
\csname @ifundefined\endcsname{endmcitethebibliography}  {\let\endmcitethebibliography\endthebibliography}{}

%\mrc{Title page for supporting info}
\newpage

%\vspace*{0.5cm}

\begin{center}
    \Large Supporting Information for
    
    \vspace{0.5cm}
    \LARGE \textbf{Tunable magnon emission from a
nano-optomagnet}
    
    \vspace{0.5cm}
    \large
    A. Duvakina, V. Karakhanyan, M. Xu, M. Suarez, A.J.M. Deenen, M. Raschetti, A. Mucchietto, T. Grosjean*, and D. Grundler*
    %\textsuperscript{*}
    
    \vspace{0.5cm}
 % Corresponding author: \href{mailto:thierry.grosjean@univ-fcomte.fr}{thierry.grosjean@univ-fcomte.fr}
 % \href{mailto:dirk.grundler@epfl.ch}{dirk.grundler@epfl.ch}
\end{center}

\vspace{1.5cm}

\noindent

\newpage

%\section*{Section S3. Figures S3.1–S3.5}
\section*{Section S1}
%\addcontentsline{toc}{section}{Section S3. Figures S3.1–S3.5}
The laser beam configuration with parameters $\rho = 450$ nm and $\beta = -30^\circ$ is referred to as Position 1. This configuration serves as the primary measurement condition used in the main experiments. It was selected based on position-dependent BLS measurements (see Figure S2), as it provided a strong and reliable signal while being sufficiently distant from the disc center to avoid optical blockage by the gold disc. The angle $\beta = -30^\circ$ was selected because, for Nanodisc I, this direction yielded the highest amplitude in the measured BLS signal.
Additionally, from these measurements, the decay length of the signal was estimated to be $l_d$ = (556 ± 128) nm.

\newpage

\renewcommand{\thefigure}{S\arabic{figure}}
\setcounter{figure}{0}
\begin{figure}[H]
     \centering
     \includegraphics[width=0.4\linewidth]{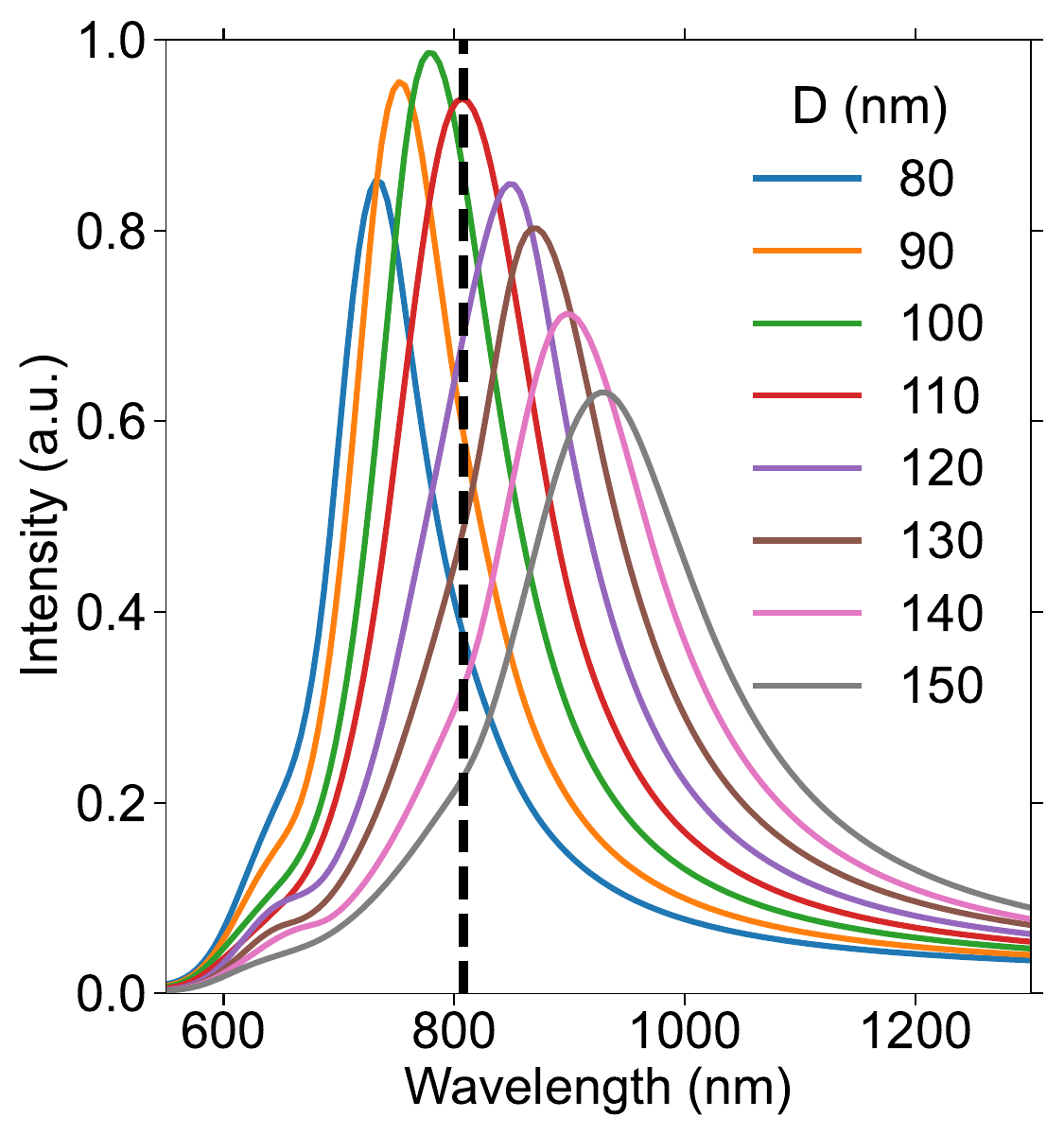}
    \caption{Simulated optical resonance spectra of plasmonic Au nanodisks positioned on a 113-nm-thick YIG film atop a GGG substrate, for antenna diameters ranging from 80 to 150 nm. The vertical dashed line indicates the IR excitation wavelength (808 nm). Simulations were carried out using the finite-difference time-domain (FDTD) method. YIG and GGG were modeled as isotropic, dispersion-free dielectric materials with refractive indices of 2.2 and 1.5, respectively. Gold dispersion was described using a Drude–Lorentz model. A non-uniform spatial mesh was used, with grid size ranging from 2 nm near the nanodisk to 20 nm at the simulation boundaries. }
    \label{fig:spectra_plot}
\end{figure}

\newpage

\begin{figure}[H]
     \centering
     \includegraphics[width=0.5\linewidth]{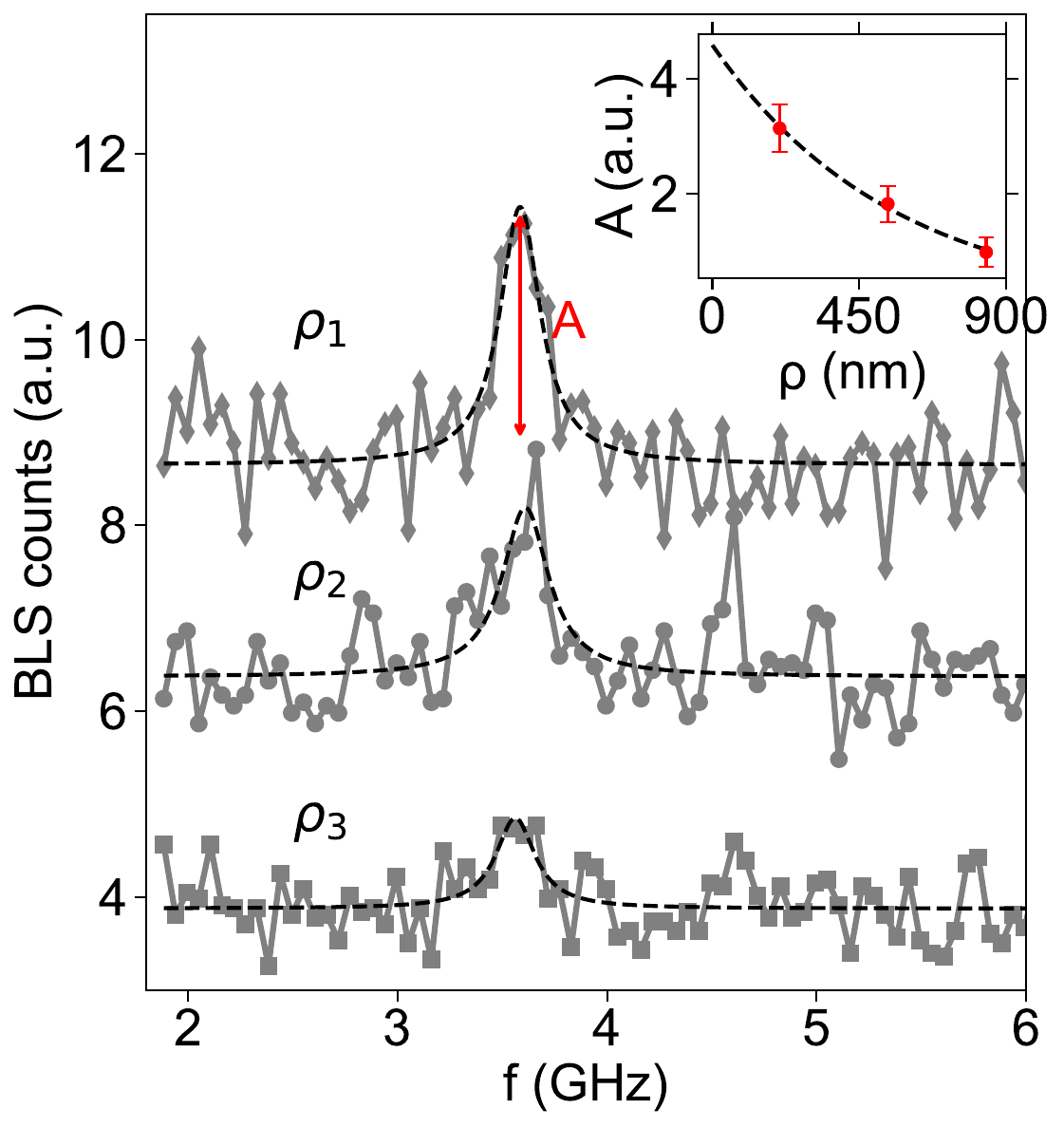}
    \caption{Position-Dependent BLS Spectra and Estimation of Decay Length. Three BLS spectra were recorded for a modulation frequency of 3.6 GHz at spatial positions $\rho_1 = 450$ nm, $\rho_2 = 537$ nm, and $\rho_3 = 838$ nm from the center of the plasmonic nanoantenna, corresponding to angles $\beta = -75^\circ, -57^\circ, -52^\circ$, respectively. The variation in angle arose from systematic calibration offsets in the scanning stage which were corrected. Each spectrum was fitted with a Lorentzian function to extract the peak amplitude. The amplitudes were then plotted as a function of $\rho$ as shown in the inset of the figure. An exponential decay function (dashed line) was fitted, and the decay constant was extracted yielding a decay length of $l_d$ = (556 ± 128) nm. Considering the anisotropic magnon dispersion relations in YIG for 20 mT we expected the excitation of at least three magnon modes at 3.6 GHz with different wave vectors resulting in wavelengths of 326, 408 and 981 nm. All these wavelengths were detectable by microfocus BLS. Hence, the observed decay was most likely caused not only by the relaxation of the individual modes but also by the dephasing of spin-precessional motion between the magnon modes along the propagation path.  The individual mode might have a larger decay (relaxation) length than the extracted value.
}
    %\label{fig:fftmag_micromag}
\end{figure}

\newpage

%\section*{Figure S3}

\begin{figure}[H]
     \centering
     \includegraphics[width=0.5\linewidth]{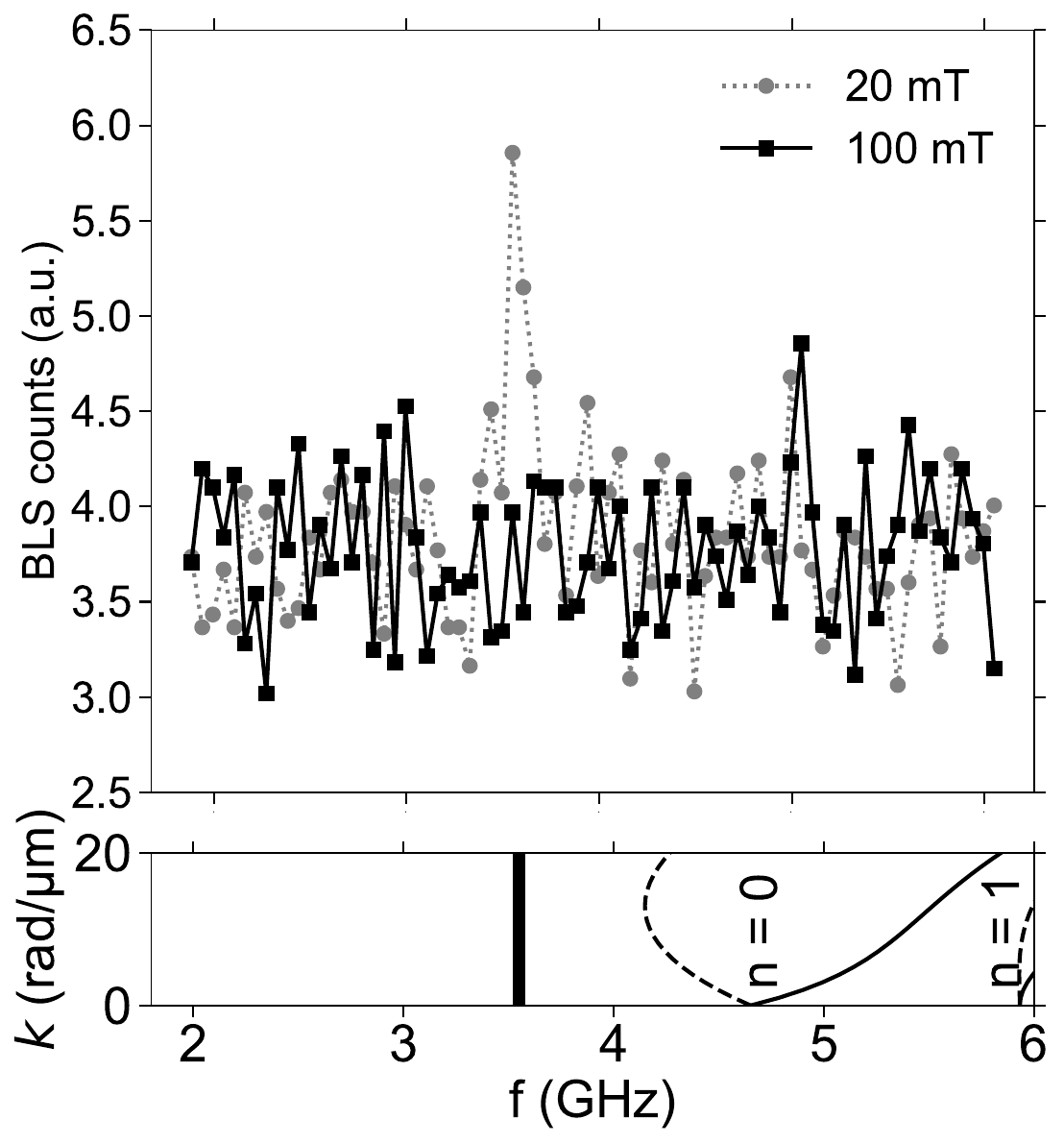}
    \caption{BLS spectra for Nanodisc I ($\rho = 200$ nm, $\beta = 165^\circ$) under $\mu_0H$ = 20 mT (grey dotted line) and 100 mT (black solid line), showing magnon signal enhancement at the IR laser modulation frequency (3.6 GHz, red) at 20 mT. No enhancement is observed at 100 mT, where the magnon band minimum lies above 3.6 GHz. 
    Bottom inset: Calculated spin-wave dispersion for a 113 nm-thick YIG film in Damon-Eshbach (solid) and backward-volume (dashed) geometries for n = 0, 1 PSSW modes under 100 mT in-plane field. Vertical line marks 3.6 GHz.}
    \label{fig:100mT_data}
\end{figure}

%\section*{Figure S4}
\newpage

\begin{figure}[H]
     \centering
     \includegraphics[width=1\linewidth]{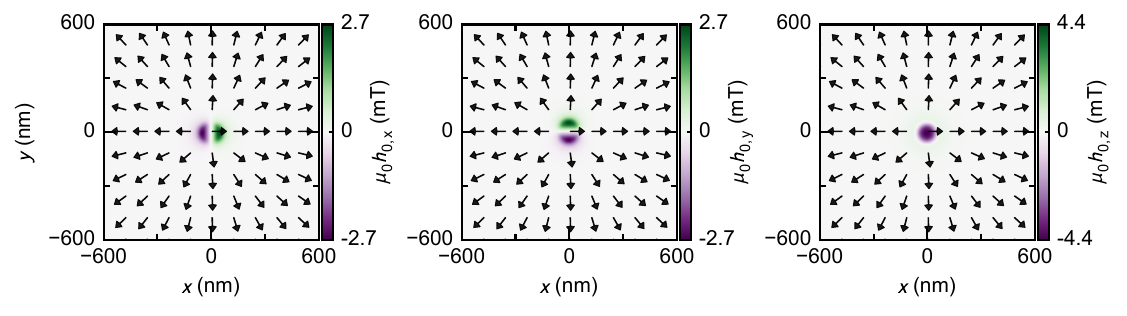}
    \caption{Numerically computed distribution of the components of the optomagnetic field $h_{0}(r)$ inside the YIG. The field shown is taken at the second layer from the Au/YIG interface. The arrows indicate the direction of the in-plane field components.}
    \label{fig:field_micromag}
\end{figure}

%\section*{Figure S5}

\begin{figure}[H]
     \centering
     \includegraphics[width=0.8\linewidth]{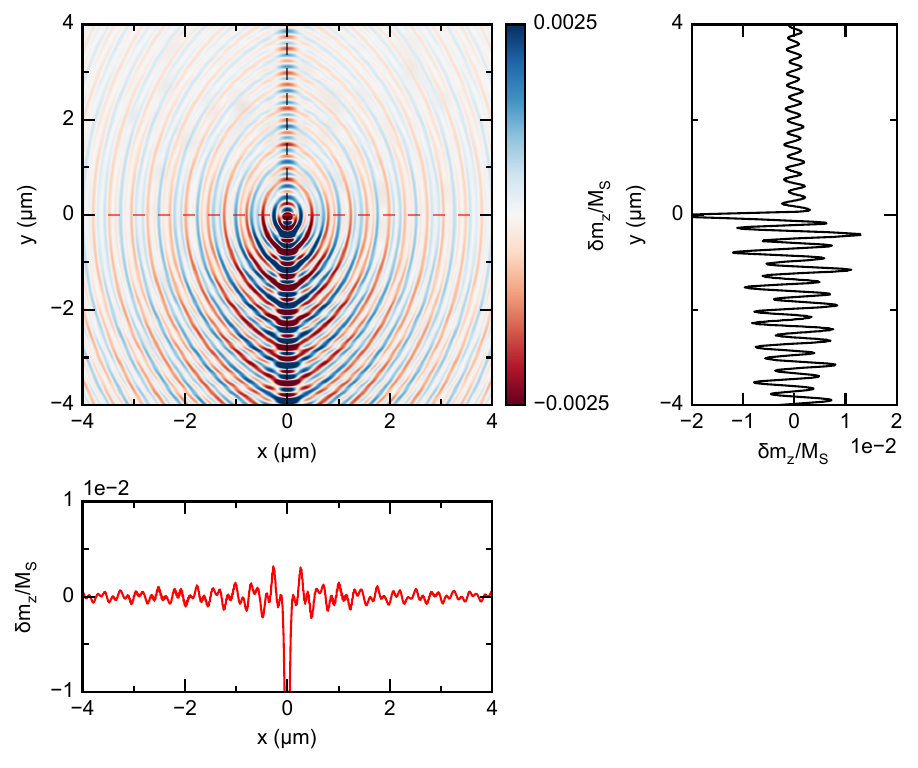}
    \caption{Snapshot in time of the simulated dynamic magnetization $\delta m_{\rm z}$ in the steady-state under continuous excitation from the field profile of Figure \ref{fig:field_micromag} at 3.6 GHz following the protocol described in the main text. The data shown is taken at at the second layer from the Au/YIG interface. Line cuts show the dynamic magnetization along the $x-$ and $y-$direction.}
    \label{fig:mag_micromag}
\end{figure}
%\section*{Figure S6}

\newpage
\begin{figure}[H]
     \centering
     \includegraphics[width=0.8\linewidth]{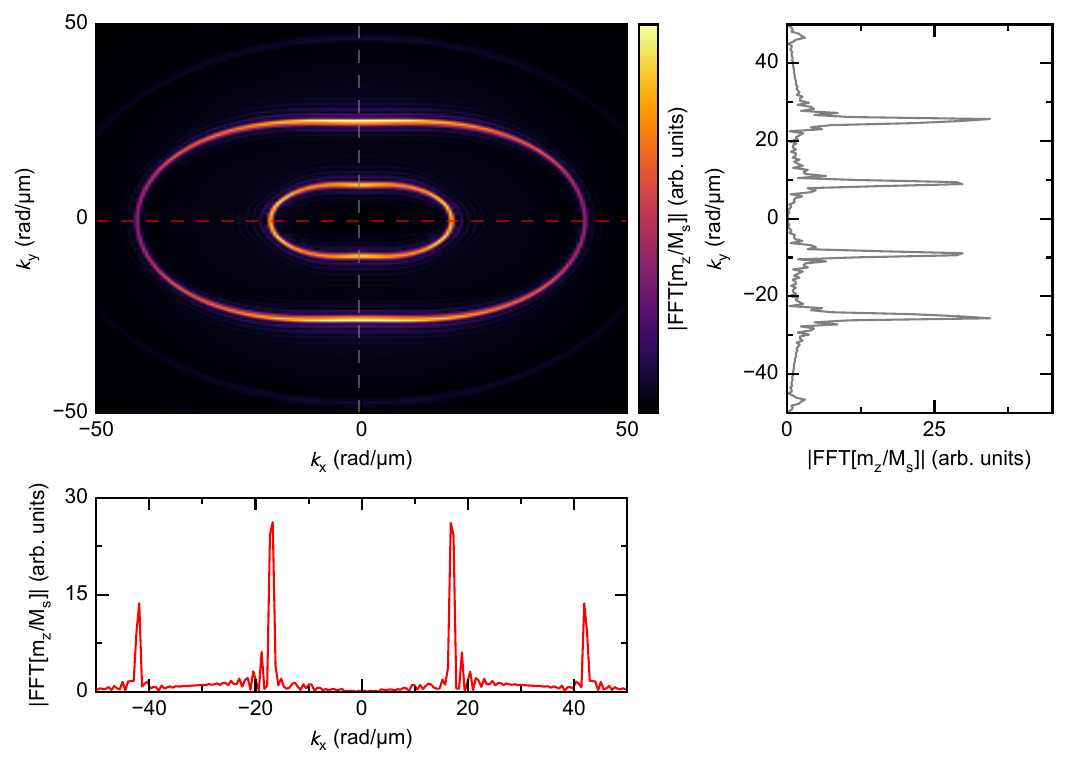}
    \caption{Norm of the complex magnetization obtained via a fast Fourier transform along the in-plane spatial coordinates of the dynamic magnetization in the steady-state shown in Figure \ref{fig:mag_micromag} at the second layer from the Au/YIG interface. Line cuts show the norm of the complex magnetization along $k_{\rm x}$ and $k_{\rm y}$. The contours reveal the presence of two modes: the fundamental mode and the $n=1$ PSSW mode, consistent with the Kalinikos-Slavin dispersion described in the main text.}
    %\label{fig:fftmag_micromag}
\end{figure}

%\section*{Figure S7}
\newpage
\begin{figure}[H]
     \centering
     \includegraphics[width=0.5\linewidth]{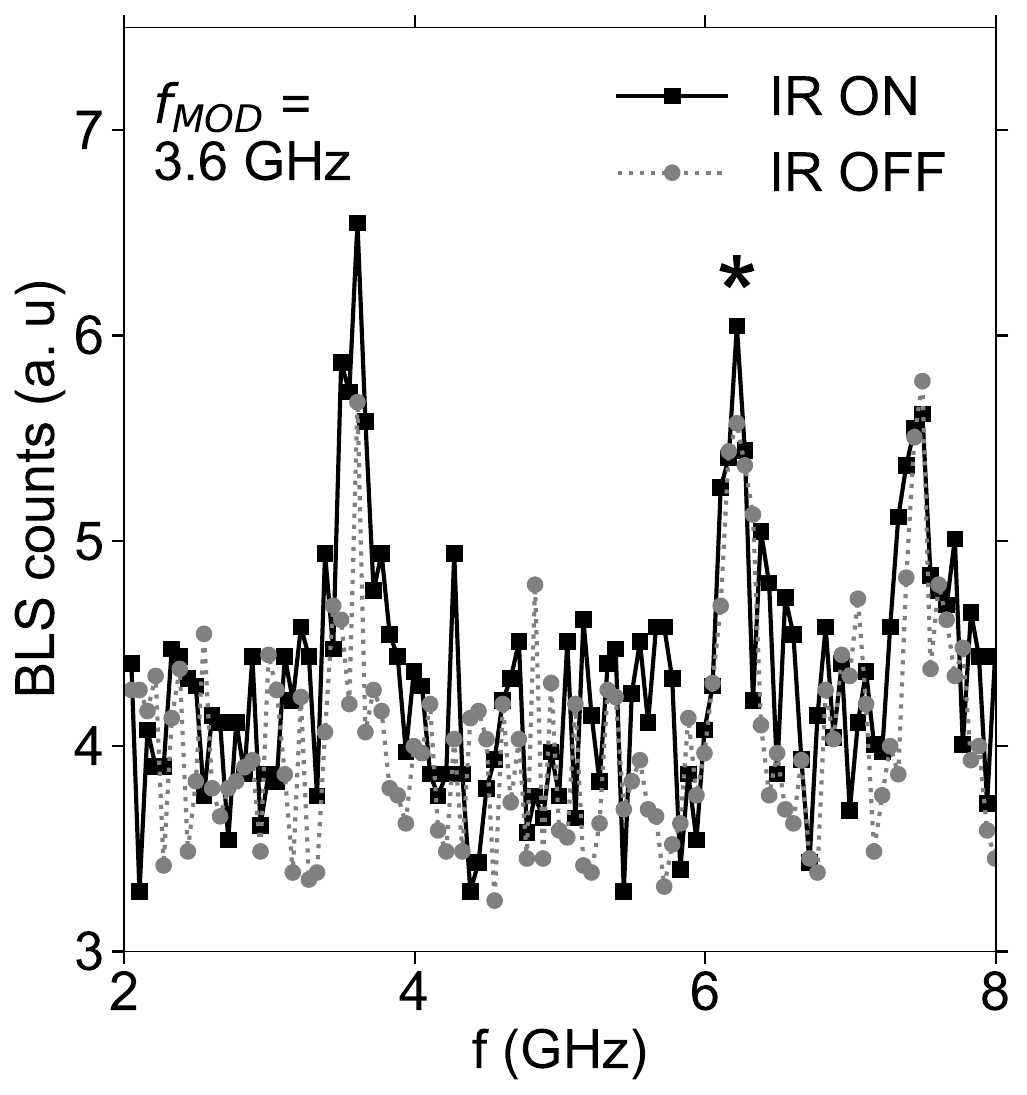}
    \caption{BLS spectra for Nanodisc I ($\rho = 450$ nm, $\beta = 90^\circ$) measured without (dashed grey line) and with (black solid line) IR excitation at modulation frequency of 3.6 GHz. The unchanged PSSW peak frequency confirms that IR excitation does not alter the spin-wave dispersion in YIG.}
    \label{fig:PSSW_peak_no_change}
\end{figure}

%\section*{References}
%\addcontentsline{toc}{section}{References}

%\end{document}


\end{document}